\documentclass[reprint,amsmath,amssymb,prl,aps,showpacs,floatfix,byrevtex,superscriptaddress]{revtex4-1}
\usepackage[english]{babel}
\usepackage{amsmath}
\usepackage{amssymb}
\usepackage{amstext}
\usepackage{amsopn}
\usepackage{amsfonts}
\usepackage{amsxtra}
\usepackage{color}
\usepackage{dcolumn}
\usepackage{graphicx}
\usepackage{hyperref}
\usepackage{bm}
\newcommand{\icm}{\ensuremath{~\textrm{cm}^{-1}}}

\begin{document}

\title{Coexistence of Clean- and Dirty-limit Superconductivity in LiFeAs}
\author{Y. M. Dai}
\email[]{ymdai@lanl.gov}
\affiliation{Condensed Matter Physics and Materials Science Department, Brookhaven National Laboratory,
 Upton, New York 11973, USA}
\author{H. Miao}
\author{L. Y. Xing}
\author{X. C. Wang}
\affiliation{Beijing National Laboratory for Condensed Matter Physics, Institute of Physics, Chinese Academy of Sciences, P.O. Box 603, Beijing 100190, China}
\author{C. Q. Jin}
\affiliation{Beijing National Laboratory for Condensed Matter Physics, Institute of Physics, Chinese Academy of Sciences, P.O. Box 603, Beijing 100190, China}
\affiliation{Collaborative Innovation Center of Quantum Matter, Beijing, China}
\author{H. Ding}
\affiliation{Beijing National Laboratory for Condensed Matter Physics, Institute of Physics, Chinese Academy of Sciences, P.O. Box 603, Beijing 100190, China}
\affiliation{Collaborative Innovation Center of Quantum Matter, Beijing, China}
\author{C. C. Homes}
\email[]{homes@bnl.gov}
\affiliation{Condensed Matter Physics and Materials Science Department, Brookhaven National Laboratory,
 Upton, New York 11973, USA}

\date{\today}

%
%

\begin{abstract}
The optical properties of LiFeAs with $T_c \simeq$~18~K have been determined in the normal and superconducting states. The superposition of two Drude components yields a good description of the low-frequency optical response in the normal state. Below $T_c$, the optical conductivity reveals two isotropic superconducting gaps with $\Delta_{1} \simeq 2.9$~$\pm$~0.2~meV and $\Delta_{2} \simeq 5.5$~$\pm$~0.4~meV. A comparison between the superconducting-state Mattis-Bardeen and the normal-state Drude components, in combination with a spectral weight analysis, indicates that the spectral weight associated with a band which has a very small scattering rate is fully transferred to the superfluid weight upon the superconducting condensate. These observations provide clear evidence for the coexistence of clean- and dirty-limit superconductivity in LiFeAs.
\end{abstract}


\pacs{78.20.-e, 74.25.Gz, 74.70.Xa}

\maketitle

%
%

Iron-based superconductors (FeSCs) are multiband materials with multiple superconducting (SC) gaps opening on
different Fermi surfaces in the superconducting state~\cite{Ding2008}. Understanding the properties of the SC gaps
is an essential step towards describing the pairing mechanism.  In FeSCs, superconductivity is generally achieved by
suppressing the magnetic and structural transitions in the parent compounds~\cite{Rotter2008a} through chemical
substitutions~\cite{Rotter2008,Sefat2008} which can introduce disorder.  Strong disorder, in
particular in-plane disorder, has been demonstrated to induce sub-gap absorption or pair-breaking effects
in FeSCs~\cite{Bang2009,Lobo2010,Teague2011}.  As a result, the spectroscopic features of the SC gaps, as
well as the values for $2\Delta/k_{\rm B}T_{c}$ may be affected by excess impurity scattering~\cite{Bang2009,Teague2011}
in doped materials.  Furthermore, the overlap or interaction between superconductivity and the magnetic order may
also complicate the measurement and analysis of the SC gaps in the underdoped regime.

LiFeAs presents an ideal system to clarify the properties of the SC gaps, as it is structurally simple and exhibits superconductivity with a relatively high critical temperature $T_c \simeq 18$~K in its stoichiometric form~\cite{Wang2008,Tapp2008}.  In the absence of disorder caused by chemical substitution, the nature of the SC gaps may be unambiguously determined by spectroscopic techniques. In addition, LiFeAs shows neither magnetic nor structural transitions~\cite{Chu2009b,Pratt2009,Qureshi2012}, so that the superconducting properties are not affected by the coexistence or interaction with other ordered states.

Recent studies on LiFeAs using angle-resolved photoemission spectroscopy (ARPES)~\cite{Umezawa2012,Borisenko2012} and scanning tunneling microscopy (STM)~\cite{Chi2012,Allan2012} have revealed nodeless SC gaps with values in good agreement with each other: $\Delta_{s}$ = 2.5-2.8~meV and $\Delta_{l}$ = 5.0-6.0~meV.  The presence of a large gap with $2\Delta_{l}/k_{\rm B}T_{c} \gtrsim 6$ places this material, at least partially, in the strong-coupling limit. Having consistently established the properties of the SC gaps by surface-sensitive techniques~\cite{Umezawa2012,Borisenko2012,Chi2012,Allan2012}, it is of the utmost importance to compare these results with bulk-sensitive probes. The bulk values of the SC gaps derived from specific heat measurements ($\Delta_{s} = 1.2$~meV and $\Delta_{l} = 2.6$~meV) are only half of the values determined by ARPES and STM, placing LiFeAs entirely in the weak-coupling limit~\cite{Stockert2011,Jang2012}. The specific heat by Wei \emph{et al.} revealed an even smaller SC gap on the order of 0.7~meV~\cite{Wei2010}. An optical study on LiFeAs by Min \emph{et al.}~\cite{Min2013} reported two isotropic gaps with values larger than the ones from specific heat studies, yet still much smaller than ARPES and STM measurements; on the other hand, Lobo \emph{et al.}~\cite{Lobo2015} observed no clear-cut signature of the SC gap from their recent optical data, which they attribute to clean-limit superconductivity in LiFeAs. However, the existing optical data on LiFeAs seem to suffer from surface contamination due to the extremely air-sensitive nature of this compound, as evidenced by the unexpected noise or kinks in the reflectivity spectra accompanied by the suppression or smearing of the phonon features at 240 and 270~\icm. To resolve the existing contradictions, further experimental, especially optical, investigations into the SC gaps in LiFeAs crystals that are free of surface contamination, is indispensable.

In this Letter, we have obtained the reflectivity of LiFeAs which is characterized by sharp phonon lineshapes and a lack of any anomalous features, indicating the absence of surface contamination. We provide clear optical evidence for two nodeless SC gaps with values of $\Delta_{1} \simeq 2.9$~$\pm$~0.2~meV and $\Delta_{2} \simeq 5.5$~$\pm$~0.4~meV, consistent with ARPES and STM measurements. By comparing the superconducting-state Mattis-Bardeen with the normal-state Drude components, we find that a band with very small scattering rate disappears from the finite-frequency optical conductivity upon the formation of a superconducting condensate. A spectral weight analysis indicates that the spectral weight lost at finite frequency due to the formation of the superconducting condensate is fully recovered in the superfluid weight. Our experimental results suggest that superconducting bands in both the clean- and dirty-limit coexist in LiFeAs.

%
%

\begin{figure}[tb]
\includegraphics[width=0.85\columnwidth]{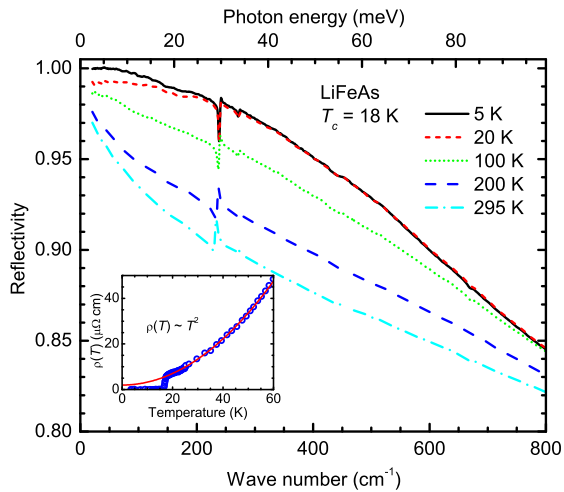}
\caption{ (color online) Far-infrared reflectivity of LiFeAs at several temperatures above and below $T_c$.
Inset: The DC resistivity as a function of temperature $\rho(T)$ in the low-temperature region (circles);
the solid curve is the fit to $\rho(T) = \rho_{0} + AT^{2}$.}
\label{fig:ref}
\end{figure}

High-quality LiFeAs single crystals were grown by a self-flux method~\cite{Xing2014}. The $T$-dependent
DC resistivity $\rho(T)$ of LiFeAs, as shown in the inset of Fig.~\ref{fig:ref}, is characterized by a sharp
superconducting transition at $T_c \simeq 18$~K.  In the normal state $\rho(T)$ follows a quadratic $T$ dependence,
$\rho(T) = \rho_0 + A T^{2}$  ($\rho_{0}$ is the residual resistivity) in the low-temperature region expected for a Fermi liquid, consistent
with previous transport studies~\cite{Heyer2011,Albenque2012}.  The residual resistivity of our crystal
($\rho_0 \approx 1.45$~$\mu\Omega$~cm) is quite small, leading to a very large
residual-resistivity ratio RRR = $\rho(300\,{\rm K})/\rho_0 \approx 200$. This indicates that
the density of impurities or defects in LiFeAs is extremely low.

%
%

Figure~\ref{fig:ref} shows the in-plane reflectivity $R(\omega)$ of LiFeAs in the far-infrared region at several different temperatures. The experimental details about the $R(\omega)$ measurements are described in the supplementary material.
%
%
In the normal state, $R(\omega)$ approaches to unity at zero frequency and increases with decreasing temperature
in the far-infrared region, indicating a metallic response. Below $T_c$, at 5~K, an upturn in $R(\omega)$ develops
at low frequency, which is a clear signature of the opening of a SC gap or gaps~\cite{Li2008,Kim2010,Tu2010,Dai2013a}.

The real part of the optical conductivity, $\sigma_{1}(\omega)$, which provides direct information about
the properties of the SC gaps~\cite{Li2008,Kim2010,Tu2010,Dai2013a,Charnukha2011a}, was
determined from the Kramers-Kronig analysis of the reflectivity.  Given the metallic nature of the LiFeAs
material, in the normal state the Hagen-Rubens form $1 - R(\omega) \propto \omega^{-2}$ was used for the
low-frequency extrapolation, while in the superconducting state $1 - R(\omega) \propto \omega^{4}$  was used.
For the high-frequency extrapolation, we assumed a constant reflectivity above the highest-measured frequency
up to 12.5~eV, followed by a free-electron response $R(\omega) \propto \omega^{-4}$.

\begin{figure}[tb]
\includegraphics[width=0.85\columnwidth]{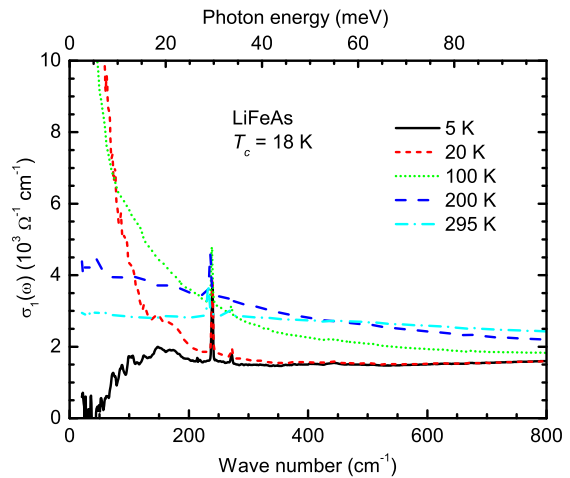}
\caption{ (color online) The real part of the optical conductivity for LiFeAs in the
far-infrared region at several temperatures above and below $T_c$.}
\label{fig:sig}
\end{figure}

Figure~\ref{fig:sig} displays $\sigma_{1}(\omega)$ for LiFeAs up to 800~\icm\ for
different temperatures above and below $T_c$.
%
%
The normal-state far-infrared $\sigma_{1}(\omega)$ exhibits a Drude-like metallic response, which
may be described as a peak centered at zero frequency where the width of the Drude response at half maximum is the value
of the quasiparticle scattering rate. As the temperature is reduced, the scattering rate decreases, resulting
in a narrowing of the Drude peak. Just above $T_c$ at 20~K, as shown by the short-dashed curve, the Drude peak
is quite narrow, suggesting a very small quasiparticle scattering rate at low temperature.  Upon entering the
superconducting state, as shown by $\sigma_{1}(\omega)$ at 5~K (solid curve), the low-frequency Drude-like
response is no longer observed, and a dramatic suppression of $\sigma_{1}(\omega)$ at low frequency sets in,
signaling the opening of the SC gaps.  The conductivity almost vanishes below $\sim$~50~\icm, suggesting the
absence of nodes in the SC gaps, consistent with ARPES~\cite{Borisenko2012,Umezawa2012} and STM~\cite{Chi2012,Allan2012},
as well as a previous optical study~\cite{Min2013}.

%
%
\begin{figure}[tb]
\includegraphics[width=0.85\columnwidth]{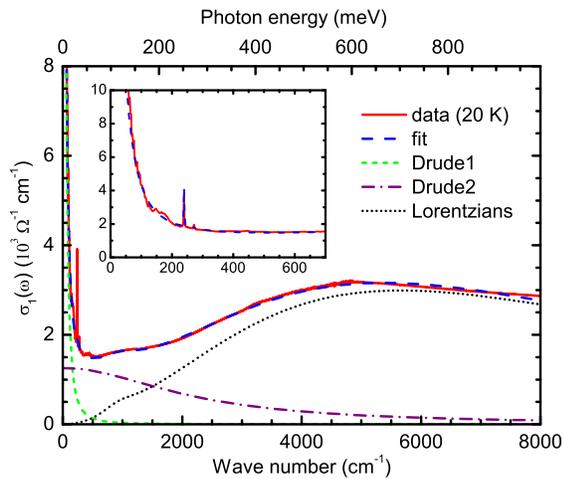}
\caption{ (color online) The solid line shows the measured $\sigma_{1}(\omega)$ of LiFeAs up
to 8000~\icm\ (1 meV) at 20~K. The long-dashed line through the data is the Drude-Lorentz fitting
result, which consists of the contributions from a narrow Drude (short-dashed line), a broad Drude
(long-dash-dot line) and a series Lorentz components (dotted line). The inset displays
$\sigma_{1}(\omega)$ (solid curve) and the fitting result (long-dashed line) in the low-frequency range.}
\label{fig:fit20K}
\end{figure}
The normal-state $\sigma_{1}(\omega)$ of this multiband material is best described using the Drude-Lorentz model \cite{Wu2010a},
%
%
\begin{equation}
\sigma_{1}(\omega) = \frac{2\pi}{Z_{0}} \left[
   \sum_{k} \frac{\omega^{2}_{p,k}}{\tau_{k}(\omega^{2}+\tau_{k}^{-2})} +
   \sum_{j} \frac{\gamma_{j} \omega^{2} \Omega_{j}^{2}}{(\omega_{j}^{2} - \omega^{2})^{2} + \gamma_{j}^{2} \omega^{2}}
 \right],
\label{DLModel}
\end{equation}
where $Z_{0} \simeq 377\, \Omega$ is the vacuum impedance. The first term corresponds to a sum of free-carrier Drude
responses where $\omega_{p, k}$ and $1/\tau_{k}$ are the plasma frequency and scattering rate in the $k$th intraband
contribution, respectively; the second term describes a sum of Lorentz oscillators, with $\omega_{j}$, $\gamma_{j}$ and
$\Omega_{j}$ being the resonance frequency, width and strength of the $j$th vibration or bound excitation.  The solid
curve in Fig.~\ref{fig:fit20K} is the experimental $\sigma_{1}(\omega)$ at 20~K, while the long-dashed line is the
fit to the data; the fitted line consists of a narrow (coherent) Drude component with $\omega_{p,1}
\simeq$~8500$~\pm$~400~\icm\ and $1/\tau_{1} \simeq$~27~$\pm$~3~\icm\ (short-dashed line), a broad (incoherent) Drude component with
$\omega_{p,2} \simeq$~13\,000$~\pm$~500~\icm\ and $1/\tau_{2} \simeq$~2200$~\pm~$150~\icm\ (dash-dot line), along with several Lorentz oscillators
(dotted line) that describe interband transitions~\cite{Marsik2013} and infrared-active phonons. The inset of Fig.~\ref{fig:fit20K}
displays the fitting result in the far-infrared region.  This approach has been widely employed to describe the optical
response of FeSCs~\cite{Tu2010,Nakajima2013a,Charnukha2013,Dai2013,Nakajima2014}.

\begin{figure}[tb]
\includegraphics[width=0.85\columnwidth]{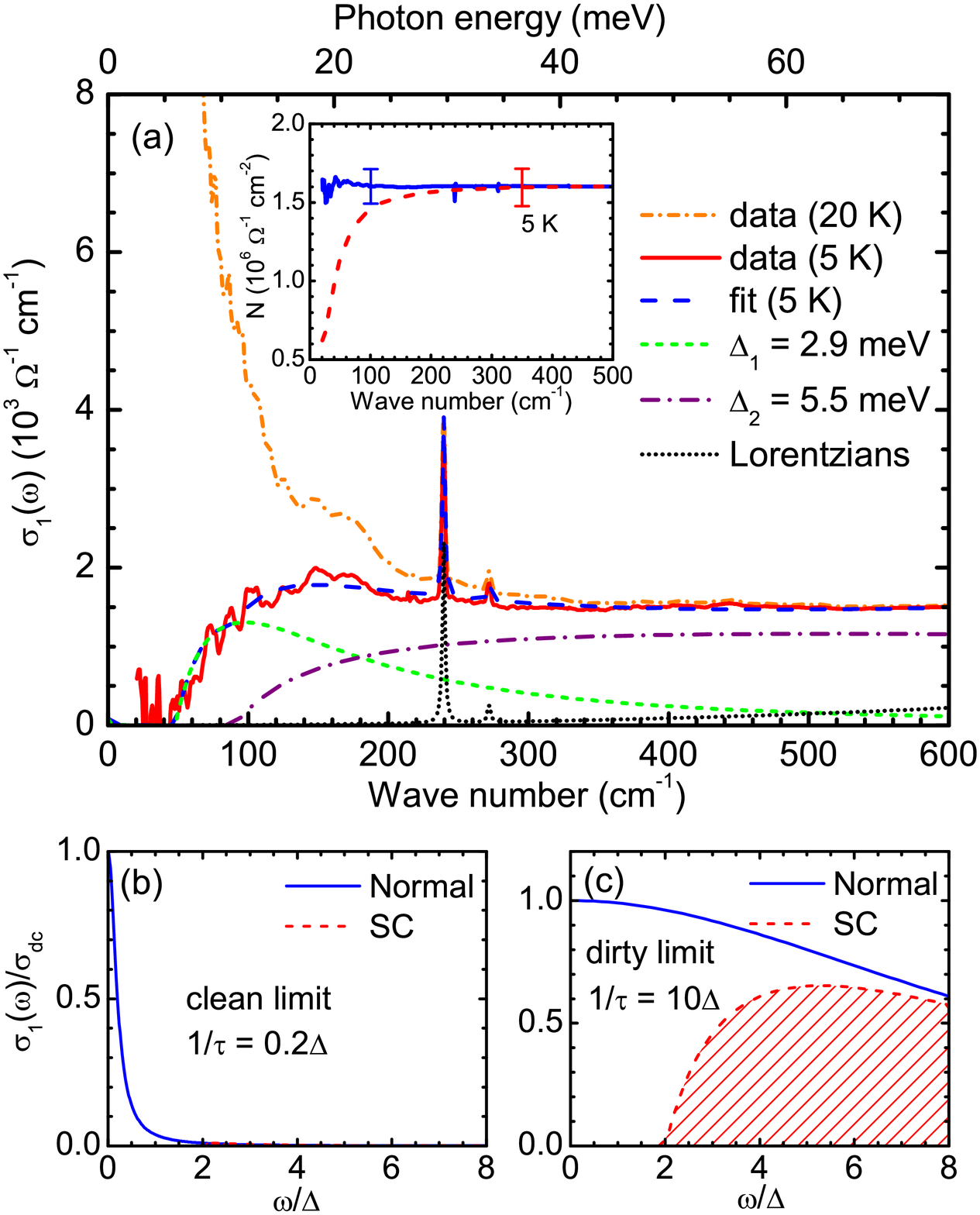}
\caption{ (color online) (a) The short dash-dot curve and the solid curve are the measured $\sigma_{1}(\omega)$
in the normal (20~K) and superconducting (5~K) states, respectively. The long dashed line thought the data at 5~K
denotes the calculated $\sigma_{1}(\omega)$ with two gaps of $\Delta_{1} \simeq$~2.9~meV (short dashed line) and
$\Delta_{2} \simeq$~5.5~meV (long dash-dot line). Inset: the superfluid weight $N_s$ calculated from the imaginary part
of the optical conductivity (solid line) and the missing area $\Delta N(\omega_{c})$ (long-dashed line), respectively.
(b) and (c) illustrate the optical conductivity in the clean- and dirty-limit case, respectively.}
\label{fig:fit5K}
\end{figure}

Having modeled the normal-state optical response, we proceed to the analysis of the data below $T_{c}$. Generally, the superconducting-state $\sigma_{1}(\omega)$ is reproduced by introducing an isotropic superconducting energy gap on each of the Drude bands using a Mattis-Bardeen formalism (supplementary material). As shown in Fig.~\ref{fig:fit5K}(a), the linear superposition of two isotropic SC gaps~\cite{Wu2010a,Nakajima2010,Homes2010,Dai2013a} with the same (unchanged) Lorentz terms from the normal state yields a very good fit to the experimental data at 5~K.
%
%
The gap values determined from the fit are $\Delta_{1} \simeq 2.9$~$\pm$~0.2~meV and $\Delta_{2} \simeq 5.5$~$\pm$~0.4~meV, respectively, in good
agreement with the photoemission~\cite{Borisenko2012,Umezawa2012} and tunneling~\cite{Chi2012,Allan2012} studies; however, both are
larger than a previous optical result on the same material~\cite{Min2013}.  The ratio of $2\Delta_{1}/k_{\rm B}T_{c} \simeq 3.7$
for the small gap is consistent with the BCS weak-coupling limit of 3.5, whereas $2\Delta_{2}/k_{\rm B}T_{c} \simeq 7.1$ for
the large gap, pointing to strong-coupling superconductivity in LiFeAs. The coexistence of weak- and strong-coupling
behaviors is likely to be a common feature in FeSCs.

Although the Mattis-Bardeen approach describes the superconducting-state $\sigma_{1}(\omega)$ quite well, and gives reasonable values for the SC gaps, we notice that while the plasma frequency of the broad Mattis-Bardeen component takes the same value as the corresponding normal-state Drude term ($\omega^{\prime}_{p,2} = \omega_{p,2} \simeq$~13\,000~$\pm$~500~\icm), the plasma frequency of the narrow Mattis-Bardeen ($\omega^{\prime}_{p,1} \simeq$~4500~$\pm$~200~\icm) is much smaller than the corresponding Drude term ($\omega_{p,1} \simeq$~8500~$\pm$~400~\icm). Here we would like to point out that in our initial fit, the plasma frequencies for both Mattis-Bardeen terms adopt the values from the corresponding Drude terms ($\omega^{\prime}_{p,1} = \omega_{p,1}$ and $\omega^{\prime}_{p,2} = \omega_{p,2}$). However, in order to achieve a reasonable fit to the data, the plasma frequency for the narrow Mattis-Bardeen component $\omega^{\prime}_{p,1}$ has to be reduced. This indicates that a band disappears from the finite-frequency $\sigma_{1}(\omega)$ upon the superconducting condensate. More interestingly, the scattering rate of the narrow Mattis-Bardeen ($1/\tau^{\prime}_{1} = 128 \pm 8$~\icm) becomes larger than the normal-state Drude component ($1/\tau_{1} = 27 \pm 3$~\icm). This is unusual, since the quasiparticle scattering rate usually decreases slightly in the superconducting state. However, if we take the disappeared band into account, this behavior can be well understood by considering that the normal-state Drude component with $1/\tau_{1} = 27 \pm 3$~\icm\ indeed describes the momentum average of a band with $1/\tau^{\prime}_{1} = 128 \pm 8$~\icm\ and another band with a very small scattering rate ($1/\tau^{\prime\prime}_{1} \ll 27$~\icm) which are strongly correlated with each other. Upon the formation of a superconducting condensate, the band with $1/\tau^{\prime\prime}_{1} \ll 27$~\icm\ disappears from the finite-frequency $\sigma_{1}(\omega)$, so that only the band with $1/\tau^{\prime}_{1} = 128 \pm 8$~\icm\ can be observed in the superconducting state.

A spectral weight analysis provides clues about whether the disappeared band participates in the superconducting condensate. The spectral weight is defined as the area under $\sigma_{1}(\omega)$ over a given frequency interval,
%
%
\begin{equation}
  N(\omega_{c}) = \int_{0^+}^{\omega_{c}} \sigma_{1}(\omega)\, d\omega,
  \label{SW}
\end{equation}
where $\omega_{c}$ is a cut-off frequency. In the superconducting state, the low-frequency spectral weight is significantly suppressed due to the formation of the SC gaps. According to the Ferrell-Glover-Tinkham (FGT) sum rule~\cite{Glover1956,Tinkham1959}, the spectral weight lost at finite frequencies in $\sigma_{1}(\omega)$ due to the superconducting condensate is transferred to the superfluid weight $N_{s}$; this is precisely the superfluid density, which may be calculated from the imaginary part of the optical conductivity $\sigma_{2}(\omega)$ (see supplementary material). The spectral weight lost at finite frequencies due to superconducting condensate $\Delta N(\omega_{c})$, the so-called ``missing area'', can be determined from a simple integral,
%
%
\begin{equation}
  \Delta N(\omega_{c}) \simeq \int_{0^+}^{\omega_{c}} \sigma_{1}(\omega,\,20\,{\rm K})- \sigma_{1}(\omega,\,5\,{\rm K})\, d\omega.
  \label{SW}
\end{equation}
The FGT sum rule requires that $\Delta N(\omega_{c})$ is equal to $N_{s}$ as long as $\omega_{c}$ covers the spectrum of excitations responsible for the superconducting condensate, regardless of the details of the system. $N_{s}$ and $\Delta N(\omega_{c})$ are shown as solid and dashed curves, respectively, in the inset of Fig.~\ref{fig:fit5K}(a). $N_{s}$ and $\Delta N(\omega_{c})$ merge together above 350~\icm, suggesting that the spectral weight lost at finite frequencies in the superconducting state, including the spectral weight associated with the disappeared band, is fully captured by the superfluid weight located at zero frequency. The superfluid plasma frequency $\omega_{\text{ps}} \simeq 7822$~\icm\ is calculated from $N_s$ via $\omega^{2}_{\text{ps}} = Z_{0}N_{s}/\pi^{2}$. The penetration depth $\lambda = 1/2\pi\omega_{\text{ps}}$ is 204~$\pm$~8~nm, which is smaller than the value from previous optical studies~\cite{Min2013,Lobo2015}, but very close to the values from other techniques~\cite{Pratt2009,Inosov2010PRL,Kim2011}.

The above observations precisely reflect the optical response of a clean-limit superconducting band. A clean-limit superconductor is described as $l \gg \xi$, where $l \approx v_{F}\tau$ ($v_{F}$ denotes the Fermi velocity) is the mean free path and $\xi \approx v_{F}/\Delta$ is the coherence length~\cite{Dressel}. Hence, the clean-limit case is also given by $1/\tau \ll \Delta$ [Fig.~\ref{fig:fit5K}(b)], indicating that nearly all of the spectral weight lies below 2$\Delta$. Upon the superconducting condensate, almost all of the spectral weight collapses into the superfluid weight located at zero frequency, leaving no observable conductivity at finite frequency [hatched region in Fig.~\ref{fig:fit5K}(b)]. Therefore, a clean-limit superconducting band disappears from the finite-frequency $\sigma_{1}(\omega)$ in the superconducting state due to the superconducting condensate, and the SC gap can not be observed in the optical conductivity~\cite{Kamaras1990}. In the dirty limit [Fig.~\ref{fig:fit5K}(c)], $1/\tau \geq \Delta$, meaning that a large portion of the spectral weight lies above 2$\Delta$, which does not participate in the superconducting condensate below $T_c$. With a large part of spectral weight left at finite frequency in the superconducting state [hatched region in Fig.~\ref{fig:fit5K}(c)], the SC gap can be clearly observed and accurately modeled by the Mattis-Bardeen formalism.

In LiFeAs, at least one band is in the clean limit, while others are in the dirty limit. The dirty-limit superconducting bands allow the SC gaps to be clearly observed from the optical conductivity and properly described by the Mattis-Bardeen approach, whereas the clean-limit superconducting band transfers almost all of its spectral weight to the zero-frequency superfluid weight below $T_{c}$, thus giving rise to the disappeared band. Since the clean-limit condition is defined through a comparison of the quasiparticle scattering rate with the superconducting gap, the partial clean-limit superconductivity is expected in a \emph{multiband superconductor} with \emph{very small residual scattering rate} alongside \emph{large and small superconducting gaps}. LiFeAs satisfies the above conditions simultaneously, thus supporting the coexistence of clean- and dirty-limit superconductivity.

The presence of clean-limit superconductivity in LiFeAs is favored by a number of experimental facts:
(i) The residual resistivity of LiFeAs is very low; $\rho_{0} \approx 1.45$~$\mu \Omega$~cm for our sample,
and $\rho_{0} \approx 1.3$~$\mu\Omega$~cm in a previous transport study~\cite{Albenque2012}, resulting in
a mean free path as large as $l \approx$~2000~{\AA} at low temperature \cite{Albenque2012}.
(ii) Upper critical field studies~\cite{Lee2010,Khim2011} have determined the \emph{ab}-plane coherence
length $\xi_{ab} \approx 40$~{\AA}, which is much shorter than $l$, placing LiFeAs in the clean limit.
(iii) An investigation into the vortex behavior in LiFeAs using STM~\cite{Hanaguri2012} has revealed
a $T$-dependent vortex-core radius, direct evidence of the Kramer-Pesch effect that is expected in a
clean superconductor.
(iv) Large superconducting gaps with $\Delta \approx$ 5 and 4.2~meV have been observed by ARPES on the inner hole
and one of the electron FSs~\cite{Umezawa2012}, respectively, where the extracted quasiparticle scattering rates
are extremely small (limited by the energy resolution) in the superconducting state~\cite{Miao2015}.

%
%
To summarize, the optical properties of LiFeAs ($T_c \simeq 18$~K) have been examined above and below $T_c$. Two isotropic SC gaps with $\Delta_{1} \simeq 2.9$~$\pm$~0.2~meV and $\Delta_{2} \simeq 5.5$~$\pm$~0.4~meV are determined from the superconducting-state optical conductivity. Interestingly, a band with a very small scattering rate vanishes from the finite-frequency optical conductivity in the superconducting state, as revealed by a comparison between the superconducting-state Mattis-Bardeen and normal-state Drude components. A spectral weight analysis demonstrates that the spectral weight associated with the disappeared band is fully recovered in the superfluid weight. These observations suggest the coexistence of clean- and dirty-limit superconductivity in LiFeAs.

%
%
%
%
\begin{acknowledgements}
We thank J. P. Hu, R. P. S. M. Lobo and B. Xu for helpful discussions. Work at BNL was supported by the
U.S. Department of Energy, Office of Basic Energy Sciences, Division of Materials Sciences and
Engineering under Contract No. DE-SC0012704. Work at IOP CAS was supported by NSFC (No. 11474344 and 11220101003) and MOST (No. 2013CB921703).
\end{acknowledgements}

%
%

\end{document}